\documentclass[trackchanges]{aastex701}
\usepackage{soul}
\usepackage{graphicx}
\usepackage{hyperref}
\usepackage[table,xcdraw]{xcolor}

\graphicspath{{./}{figures/}}

\begin{document}

\title{Tracking Magnetic Topological Change in a Time-Dependent Coronal Model}


\correspondingauthor{Emily I. Mason}
\email{emason@predsci.com}

\author[0000-0002-8767-7182]{Emily~I.~Mason}
\affiliation{Predictive Science Inc., 9990 Mesa Rim Rd., Ste. 170, San Diego, CA 92121, USA} 
\email{emason@predsci.com}

\author[0000-0003-1759-4354]{Cooper~Downs}
\affiliation{Predictive Science Inc., 9990 Mesa Rim Rd., Ste. 170, San Diego, CA 92121, USA} 
\email{cdowns@predsci.com}

\author[0000-0001-9231-045X]{Roberto~Lionello}
\affiliation{Predictive Science Inc., 9990 Mesa Rim Rd., Ste. 170, San Diego, CA 92121, USA} 
\email{lionel@predsci.com}

\author[0000-0003-1662-3328]{Jon~A.~Linker}
\affiliation{Predictive Science Inc., 9990 Mesa Rim Rd., Ste. 170, San Diego, CA 92121, USA} 
\email{linkerj@predsci.com}

\author[0000-0001-7053-4081]{Viacheslav~S.~Titov}
\affiliation{Predictive Science Inc., 9990 Mesa Rim Rd., Ste. 170, San Diego, CA 92121, USA} 
\email{titovv@predsci.com}

\begin{abstract}

We apply the slip-back mapping method of \cite{Titov2009} and \cite{Lionello2020} to a thermodynamic MHD simulation to track topological changes in the magnetic field at a range of temporal cadences. The method constitutes the logical successor to a simple open-field map for a steady-state model, as it tracks changes in the open and closed fields for a time-dependent model by tracking individual magnetic elements as they advect across the map, rather than simply tracing field line connectivity from each cell. Through careful categorization of the slip-back mapping values and analysis of the flux changes, we not only effectively track the open flux but can recover the flux processed through interchange reconnection as well. The field lines involved in these processes are shown to follow lines of high squashing factor, as proposed by interchange reconnection-driven slow solar wind theory. The time-dependent model, which is scaled to solar minimum-like activity, projects that {a median value of 3.5\% of the total open flux in any given 24-hour interval} has been processed through interchange reconnection. This corresponds to a relatively high proportion of the total open flux changes over time in the heliosphere. Our results show that not only is this method a useful tool for accurately tracking topological change in time-dependent simulations, but that its inherent complexity can be visually reduced into an intuitive 2D plot that simply and effectively communicates temporal changes.

\end{abstract}


\section{Introduction}\label{sec:intro}

The solar corona is in a state of constant change, as the interplay between the magnetic fields and mechanical forces drive dynamics at every length scale and temporal cadence imaginable. Despite major progress across many domains, there are still significant open questions. These include the so-called ``open flux problem'' \citep{Lowder2017,Linker2017,Wallace2019}, wherein there is a discrepancy between the in-situ measured heliospheric open flux and that which is estimated from coronal observations. Another prominent set of questions concerns where, how much, and what types of plasma populate the solar wind from different source regions in the low corona \citep[e.g.,][]{Geiss1995,Wang1998,Antiochos2011,Owens2014,Laming2015,Bale2019,Viall2020}. These questions are challenging to close because, while we have observations of the low corona and in-situ observations, there are few if any direct observations connecting them, and many different scales, physical transitions, and processes are involved in the vast distances between them \citep{Parker1965,Gosling1999,Cranmer2005,Borovsky2008,Karimabadi2013,West2023}.

Theories like the S-web \citep{Antiochos2011,Titov2011} attempt to connect these regions and explain how plasma on field lines all over the Sun will distribute itself, accelerate, and possibly escape as it moves through the corona. The theory maps out the regions of high or low magnetic convergence, highlighting regions such as coronal holes, streamers, and null-point topologies. Visualizing the three-dimensional signatures of these regions as they evolve with time and height is important to understanding both low-coronal dynamics and solar wind development and evolution \citep{Antiochos2012}. However, connecting frameworks such as the S-web with the actual behavior of the plasma populating the magnetic framework is often not straightforward, even within the ranges for which we have observations \citep{Chitta2023,Baker2023,Wallace2025}.

One of the primary mechanisms highlighted by S-web theory is topological change, which is the main topic of this paper. There are several types of topological change, briefly summarized here. Closed-closed reconnection is the exchange of footpoint identities between two closed magnetic field lines, which results in two more closed magnetic field lines; in other words, the identity of the loops changes, while their classification (two closed field lines) does not. In open-closed (or interchange) reconnection, one open field line and one closed field line exchange footpoints, resulting in the opening of the closed field line and the closing of the open field line. Here, as with closed-closed reconnection, the identity of the loops changes (one opens, one closes), but their classification does not (one open field line, one closed field line). Open-open reconnection, on the other hand, involves ``pinch-off'' reconnection between two open field lines, resulting in one disconnected field line and one closed field line. Here, both the identity and the classification changes.  Finally, a closed loop can expand outward to large distances, dragged out by the solar wind, or as part of a CME.   From the coronal point of view, this field line can be considered open when its apex is beyond the Alfv\'en surface.  Within the context of the MHD model, it will be identified as two open field lines when the apex crosses the outer boundary.

The magnetic fields in the corona respond to the evolution of the photospheric magnetic field. Models strive to quantify this response. To that end, the advent of surface flux transport models \citep{schrijver03b,argeetal2013,Upton2024,Caplan2025} has hastened the adoption of time-dependent MHD modeling, which in turn has produced more realistic results than steady-state models \citep{Mason2023TDC,Downs2025}. These advances also provide the opportunity to revisit common methods of analysis and to develop novel tools that can extract the most useful information possible from new and improved simulations. In other words, making the most of time-dependent models requires time-dependent analysis tools.

In this paper, we present a new application of slip-back mapping (SBM), first introduced in \cite{Lionello2020}. SBM allows users to investigate magnetic flux evolution and identify some of the mechanisms by which topological change is occurring throughout the simulation, with better accuracy than previous tools. This method, which depends upon thorough knowledge of the inner boundary’s flows, is necessary due to the inner boundary’s evolution in a time-dependent model. Position is no longer sufficient information to track flux evolution; each field line moves with flows, so the net advection of the parcel must be taken into account in order to accurately trace a given field line. SBM calculates these traces with a tunable time step, allowing for large-scale analysis of topological change across the entire corona. We apply this method to the time-dependent Magnetohydrodynamic Algorithm outside A Sphere \citep[MAS,][]{Mikic1999} simulation first presented in \cite{Lionello2023GlobalCorona}. Section \ref{sec:mas} discusses the background and details of the code and the time-dependent mode. Section \ref{sec:sbm} covers the background and overview of the SBM method (more specific details of the algorithm are provided in the Appendix), and includes some notes on the interpretation of the raw method output. Section \ref{sec:results} presents the results of the investigation, focusing on open flux evolution and the aggregated topological changes in the model, while Section \ref{sec:disc} discusses the implications of the findings for our understanding of the dynamic corona, both in theory and data analysis.

\section{MAS}\label{sec:mas}

MAS solves the thermodynamic, resistive MHD equations on a nonuniform spherical mesh; these equations describe coronal heating, thermal conduction parallel to the magnetic field, and radiative losses. It uses a semi-implicit time-stepping algorithm, and covers the global corona and solar wind  \citep{,Mikic1999,Lionello2009,Riley2011}.    The thermodynamic MHD approach allows the plasma density and temperature to be computed with sufficient accuracy to forward-model EUV and soft X-ray emission and other remote sensing observables \citep{Lionello2009,Downs2013,Mikic2018PredictingEclipse,linkeretal2021}.  The coronal domain extends to 30\,$R_\odot$, where outflows are typically super-Alfvénic (simplifying the outer boundary conditions).  A wave-turbulence-driven (WTD) approach is applied for coronal heating and solar wind acceleration to model the large-scale solar wind properties \citep{Downs2016,Mikic2018PredictingEclipse}.  MAS has produced state-of-the-art solutions of the corona for case studies of its structure and connectivity \citep{Mikic2018PredictingEclipse,telloni22,antonucci23,Downs2025}, coronal mass ejections \citep{linkeretal2003,Lionello2013b,Torok2018,downs2021}, and the inner heliosphere in general \citep{Riley2011,Riley2019}.

Recently, MAS has incorporated a method for evolving $B_r$ at the inner boundary, $B_{r0}(\theta,\phi,t)$, such that it matches the values supplied by the flux transport model. The evolution is specified via the tangential electric field at the boundary, 
\begin{equation}
    E_t = \nabla_t \times \Psi \hat{r} + \nabla_t \Phi,
\end{equation}\label{eq:Et}by solving for two scalar potentials, $\Psi$ and $\Phi$ \citep{Mikic2018PredictingEclipse,Yeates2018}, the first of which controls the $B_r$ evolution. When employing flux-transport models and maps, the large-scale flows (differential rotation and meridional flows) influence the long-term field evolution; these are included in the second, non-inductive potential. The strength of the method is that all known boundary flows can be incorporated into $E_{t0}$, while ensuring that $B_{r0}(\theta,\phi,t)$ always matches the specified values \citep[as discussed in][]{Lionello2023GlobalCorona}. Freedom in the second potential can also be further adapted for the introduction of energization. {However in this case, this potential is not used to inject additional shear, helicity, and magnetic energy at large-scale neutral lines \citep[as was done in][]{Downs2025}, and as such, this model represents a minimally energized corona evolving at solar minimum.}

For this study, we use the original time-dependent simulation in MAS \citep{Lionello2023GlobalCorona,Mason2023TDC}, in which the evolving maps of the photospheric field were provided by the Lockheed Evolving Surface-Flux Assimilation Model \citep{schrijver03b}. In this case, a ``synthetic'' sun was simulated, with the evolving surface fields scaled and distributed such that they approximate the conditions near solar minimum. This model includes small- and large-scale flux emergence, flux decay, differential rotation, and meridional flows. Using fully synthetic data (created using observationally-derived statistical distributions) ameliorates many of the challenges in magnetic data assimilation methods, such as far-side flux emergence and global flux balancing. The sequence of $B_r$ maps was utilized with a cadence of one hour of solar time per map; there were 721 maps in total, spanning $\sim$30 days of physical evolution. 

\section{Slip-back Mapping}\label{sec:sbm}

\subsection{Anatomy of a Mapping}
The algorithm behind the slip-back mapping (SBM) method was originally outlined in \cite{Titov2009} and expanded in \cite{Lionello2020}. Specific details on the implementation of SBM for this time-dependent run can be found in the appendix of this paper, but we describe the method in general terms here.

To begin, consider two times in a time-dependent simulation, which we designate $t_0$ and $t_1$. Two slip surface locations are designated from which the mappings begin or end: in this study, the inner and outer slip surfaces are located at $r=1.01$ and $r=19 \,R_\odot$, respectively.

\begin{figure}
    \centering
    \includegraphics[width=0.75\linewidth]{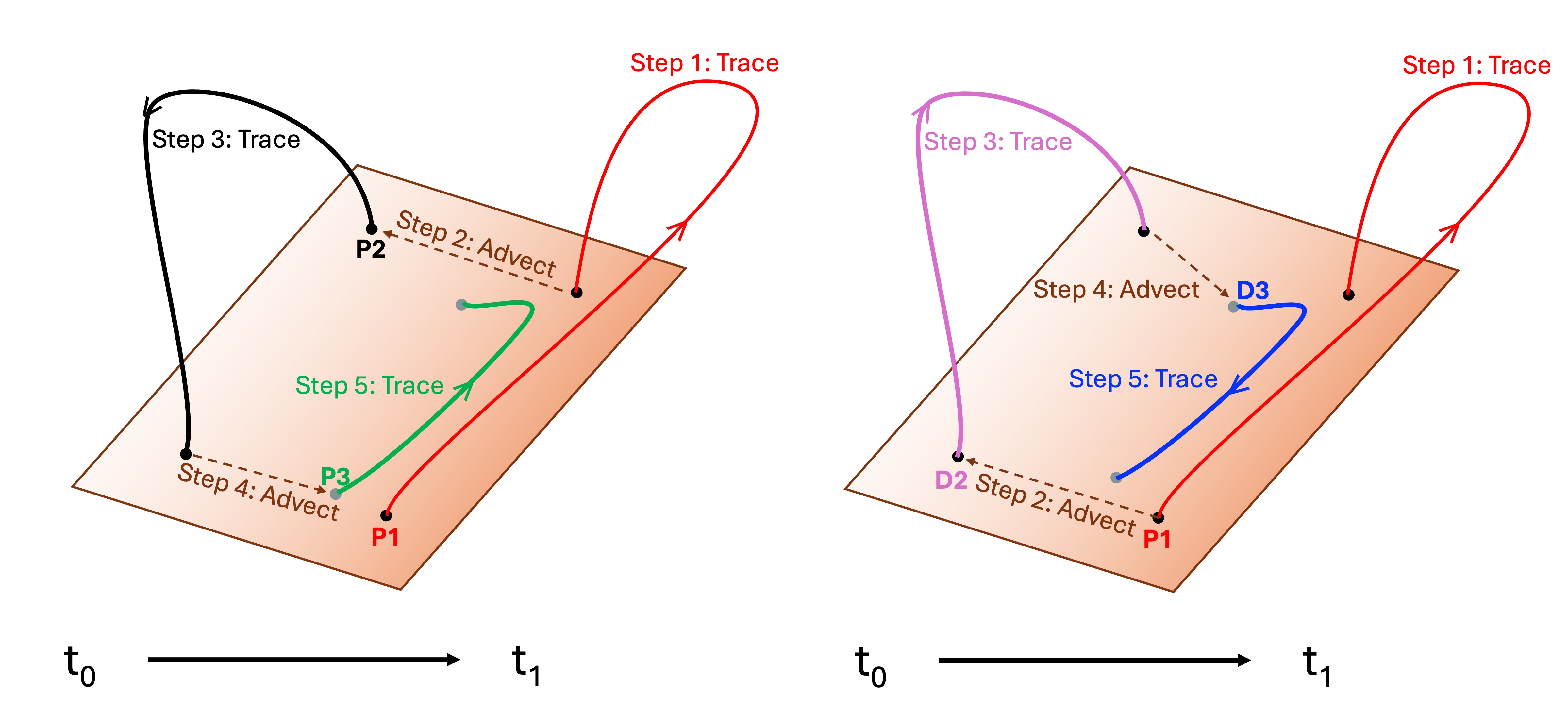}
    \caption{Diagram showing the algorithm for both the primary (a) and dual (b) mapping methods detailed here. {Each step of the mapping process is outlined in the figure, with steps 1, 3, and 5 being field line traces and 2 and 4 being advections back or forward in time between $t_0$ and $t_1$.} The footpoints whose identity determine the SBM code are marked with P (primary) or D (dual) and a number, and are shown in the same color as the field lines shown in the next panel.}
    \label{fig:mapping}
\end{figure}

To begin a mapping, the intersection of a magnetic field line with the slip surface (P1) is selected at $t_1$, and the field line is traced to its opposite end. The other end may either be located on the same slip surface -- for a closed or disconnected loop (depending upon whether the mapping begins from the inner or outer slip surface, respectively) -- or on the opposite slip surface for an open field line. This is step 1 as shown in Figure~\ref{fig:mapping}. 

After this first step, there is a divergence in the algorithm for the primary (Fig.~\ref{fig:mapping}a) and dual (Fig.~\ref{fig:mapping}b) mappings. For the primary mapping, the opposite footpoint is advected backward from $t_1$ to $t_0$ using the net flow in the plane of the slip surface, termed $v_{tfl}$. For the dual mapping, it is P1 which is advected backward in time. In either case, this is marked as step 2 in the figure. Again, the field line from P2 (or D2) is traced in step 3, and the footpoint is advected forward in time in step 4 to find P3 (or D3) at $t_1$. This point is then used to trace the final field line, completing the process in step 5.

\begin{figure}
    \centering
    \includegraphics[width=0.75\linewidth]{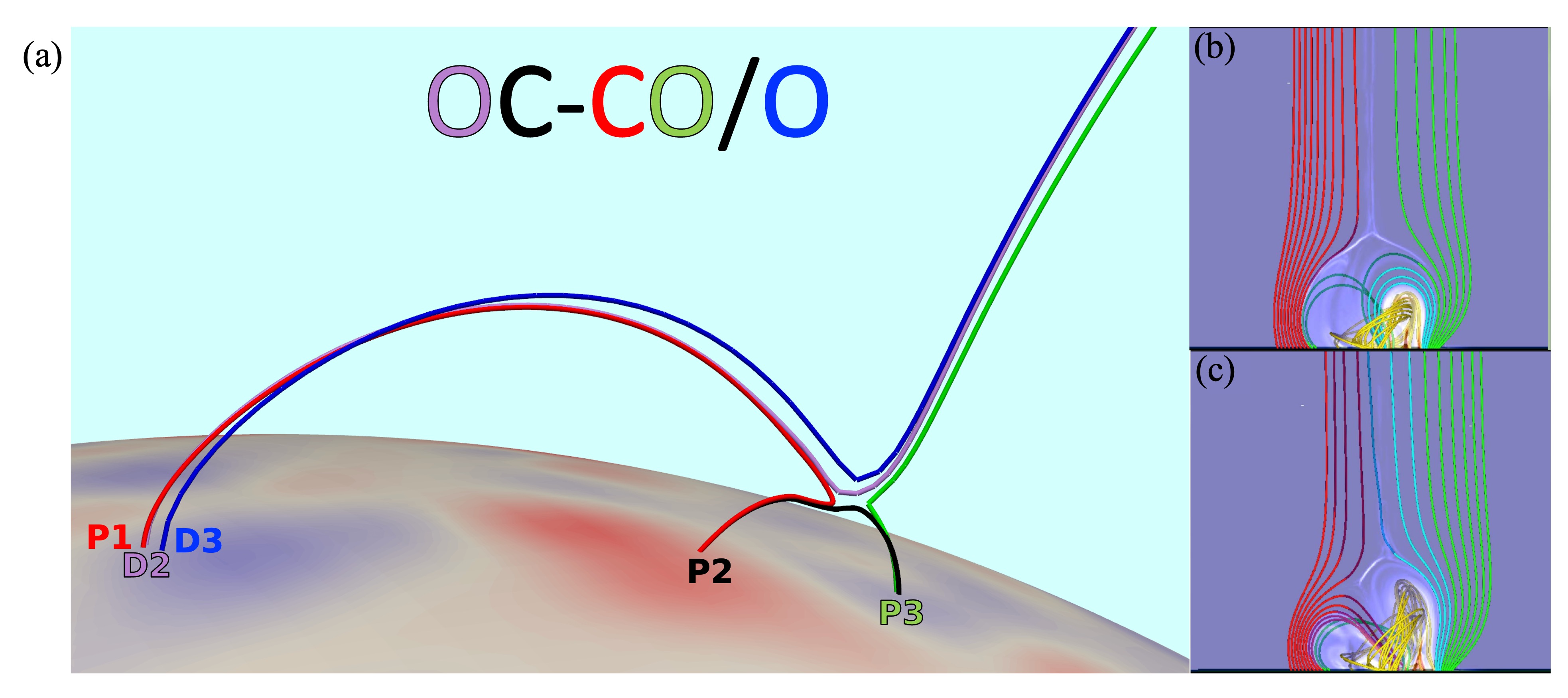}
    \caption{(a) Example plot of an interchange reconnection code, with labeled footpoints. P2 and D2 are located in the past ($t_0$), while P1, P3, and D3 are in the present ($t_1$), but all are co-plotted for context. Please note that some footpoints are very close to each other, making it difficult to distinguish between them. (b) and (c) illustrate interchange reconnection in an MHD simulation, where the open red and closed cyan field lines in (b) reconnect with each other through the current sheet (shown in white) to become the closed red and open cyan field lines (reproduced with permission from Fig. 7 of \cite{Wyper2018}).}
    \label{fig:occoo}
\end{figure}

Each theta-phi location on the simulation grid is then subjected to this process to build up the full map, and the connection values of each point (P1, P2, P3, D2, D3) are recorded. We use a value of 0 for a closed field line, 1 for an open field line, and 2 for a disconnected field line. These values are given in the order D2 P2 P1 P3 D3, resulting in a 5-digit value in base-3.  P2 and D2 are from $t_0$, while the other three are from $t_1$, hence the order. This 5-digit value is converted to base 10 and saved as the SBM value for that pixel, resulting in a final 2D theta-phi map for a given time range and cadence. When discussing specific codes, we refer to them in the same order but by their mappings, such as {OC-CO/O}.

As a simple example, we show SBM code {OC-CO/O} in Figure~\ref{fig:occoo}a. The traced field lines are colored using the same convention as the footpoints are labeled in Figure~\ref{fig:mapping}ab; namely, P1 in red, P2 in black, P3 in green, D2 in purple, and D3 in blue. This scenario, which carries the value {109} in the SBM map, represents an interchange reconnection case. Here, the field lines at $t_0$ are {open and closed} respectively (the {OC} at the beginning of the code), while both the primary and dual mappings return field lines at $t_1$ which are {closed and open} (both return {CO}, indicating the connectivity has swapped between the two field lines). Order matters in the code; OC-CO/O, for example, is not the same as OC-OO/C. The first indicates that the primary mapping indicates interchange reconnection and the dual mapping has identified an adjacent field line, while the second indicates the reverse {(note: neither is the code mapped in Fig. \ref{fig:occoo})}. A simple example of interchange reconnection that is directly analogous to (a) is shown in Fig.~\ref{fig:occoo}bc. Reconnection in particular can be challenging to track in this way, as discussed in Section \ref{sec:results}.

\subsection{Notes on the Method}\label{sec:notes}
There are a few important aspects of the SBM process that the reader should note. Although the MAS algorithm advances the vector potential and conserves magnetic flux to machine precision, the SBM calculations use the magnetic field output, which, when interpolated to obtain values at different locations, may not exactly preserve this property. Furthermore, the existence of a discrete outer boundary (in this case, the outer slip-surface) leads to some flux leaving the domain and being lost from the system. Similarly, a closed loop whose apex is advected outward will artificially be categorized as two opening field lines. These causes are well-known limitations shared by all MHD simulations. Lastly, flux can be lost or misidentified through propagated advection errors. These can build up when the output cadence of the 3D velocity data is larger than the time step of the MHD algorithm. Optimally, the SBM would be calculated by the MHD code during each time step. However, conducting this step as part of the run itself is time-intensive, so in practice we calculate it \textit{ex post facto} using the simulation outputs.  

SBM may be performed beginning from either the inner or outer slip surface, and in fact the mappings will not have a 1:1 correspondence. This is because some cases involving disconnected flux can only be detected from the outer slip surface. For this study, all calculations were made from the inner slip surface, as this is the more relevant case for coronal investigations.

In reference to the process in general, the arrows shown in Fig.~\ref{fig:mapping} designate the direction of the \textit{trace}, not the direction of the magnetic field. In fact, the direction of the trace sometimes opposes that of the field, because the first trace (from P1 in step 1) is always outward from the slip surface. This may be with the magnetic field (if P1 is located in positive field) or against it (if P1 is located in negative field). Since the dual mapping advects P1 back in time, it must trace in the same direction, as D2 must be the same polarity as P1. By contrast, the primary mapping advects the slip-surface's point of intersection at the opposite end of the field line back in time, so that tracing must go in the opposite direction.

Finally, SBM can be calculated for any desired cadence (as long as the simulation produced data at the desired rate), and indeed the results here reflect data from two separate calculations, one at a 6-hour cadence and one at a 24-hour calculation. The latter is used only for Fig.~\ref{fig:slogq_sbm} and its associated animation. The process described above assumed that $t_0$ and $t_1$ were separated by a single time step for simplicity, but the process only changes in that there are more advections backward or forward for each intervening time step, in order to maintain an appropriate trace. 

\subsection{Scale-Based Analysis Recommendations}
Although the method relies on tracing field lines from every point along the grid and tracking those field lines backward and forward in time in accordance with the slip-surface flows, interpreting individual cases involves a thorough understanding of the process used to create the maps. SBM can certainly serve as a method by which to zero in on small-scale regions undergoing rapid topological change, and to easily illustrate the relevant reconnection mechanisms occurring. However, caution must be exercised when analyzing individual cases. The flux involved in the topological change can be of the same order of magnitude as the uncertainty in the flux measurement itself, depending upon the strength of the field at that location and the grid size of the simulation. This, coupled with the uncertainty in the mapping that can be introduced by loops shorter than the slip-surface height (discussed in Sec.~\ref{sec:notes}), can result in confusing results for the unwary user. We therefore strongly recommend binning the codes into a few categories as we have done here, and it is of the utmost importance to evaluate and understand each code before determining to which category it belongs.

\section{Results}\label{sec:results}

Raw SBM maps are arrays of several hundred unique code values, spread out across a theta-phi map in a pattern reminiscent of a cobweb. This is difficult to visualize clearly or to interpret. In order to turn these disparate individual events into an intuitive dataset, we binned the values into six main categories. These categories are defined by the physical mechanism they represent, and whether it is the primary or dual mapping that identifies the mechanism. The mechanisms are flux opening (FO), flux closing (FC), and disconnection (DC). For primary mappings, we compared the state of P1 (open or closed) to the state of P2 (which could be open, closed, or disconnected). If P1 was open and P2 was closed, the code belonged to FO; the opposite situation belonged to FC. An analogous categorization held for the dual mapping, with a comparison between P1 and D2. Disconnected codes include a disconnection value at $t_1$ that is not already in the FO or FC category, to avoid double-counting between the mechanism bins (double-counting also occurs between the dual and primary mappings, which we handle by dividing the sums of these values in half, as shown in the equations below). If either the primary or dual mapping had no change between $t_0$ and $t_1$ (i.e., both P1 and P2/D2 were open, or both were closed), the code was not included in the following statistics for that bin. 

For example, the code OC-CO/O shown in Fig.~\ref{fig:occoo}a would be discounted from the primary categorization because both P1 (at $t_1$) and P2 (at $t_0$) are closed; however, it would be included in the dual categorization as FC, since P1 is closed but D2 (at $t_0$) was open. This method, of course, includes interchange reconnection as well as more straightforward flux opening or closing, such as helmet streamer growth (where closed magnetic field lines' apexes pass outside the outer boundary of the simulation, becoming \textit{de facto} open field) or pinch-off. We discuss how we recover the interchange reconnection fluxes from the FO and FC bins below.

\begin{figure}[htbp]
    \centering
    \includegraphics[width=0.9\linewidth]{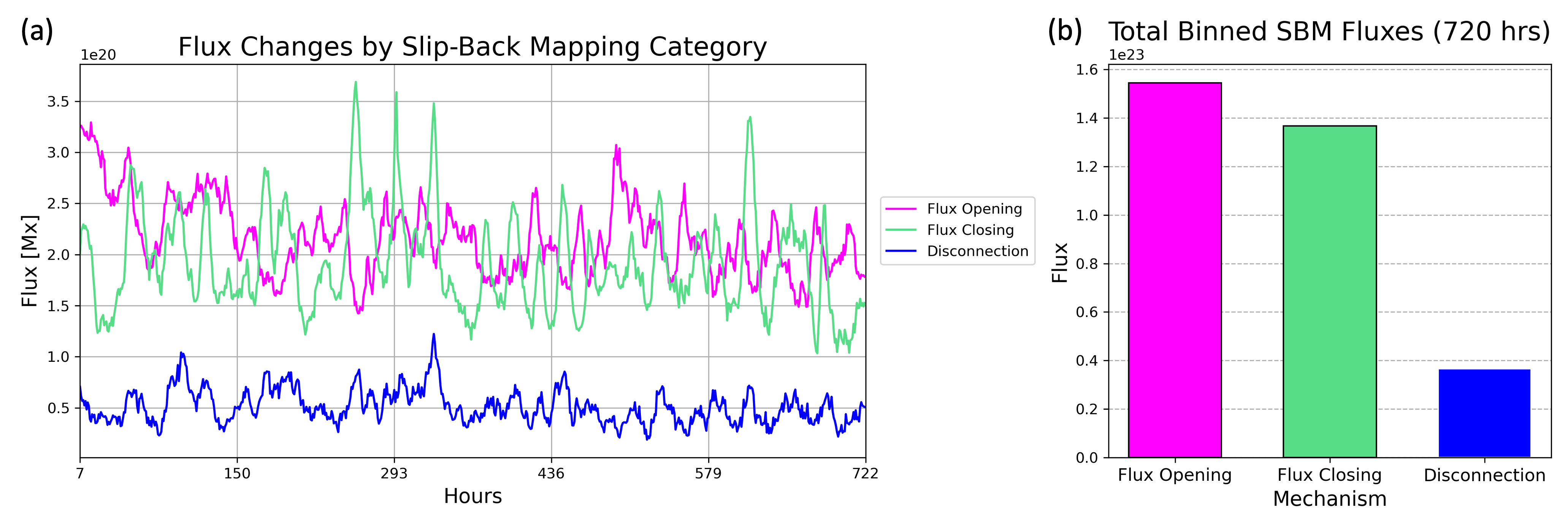}
    \caption{(a) Flux opening, closing, and disconnecting per time step, derived from both the dual and primary mappings, calculated at a 6-hour cadence for the entire TD simulation. (b) The cumulative flux processed by flux opening, closing, or disconnecting over the entire simulation using the 6-hr SBM cadence.}
    \label{fig:sbm_fluxes_multiplot}
\end{figure}

Figure~\ref{fig:sbm_fluxes_multiplot}a shows the evolution of these categories of SBM fluxes over time, using a 6-hr cadence for the mapping on the full 721-hour simulation. The FO and FC categories show much higher overall fluxes and more variability than the DC category, as expected from a SBM applied to the inner slip surface. Overall, the FO slightly exceeds the FC category, as can be seen more clearly in Figure~\ref{fig:sbm_fluxes_multiplot}b.

\begin{figure}[ht]
    \centering
    \includegraphics[width=\linewidth]{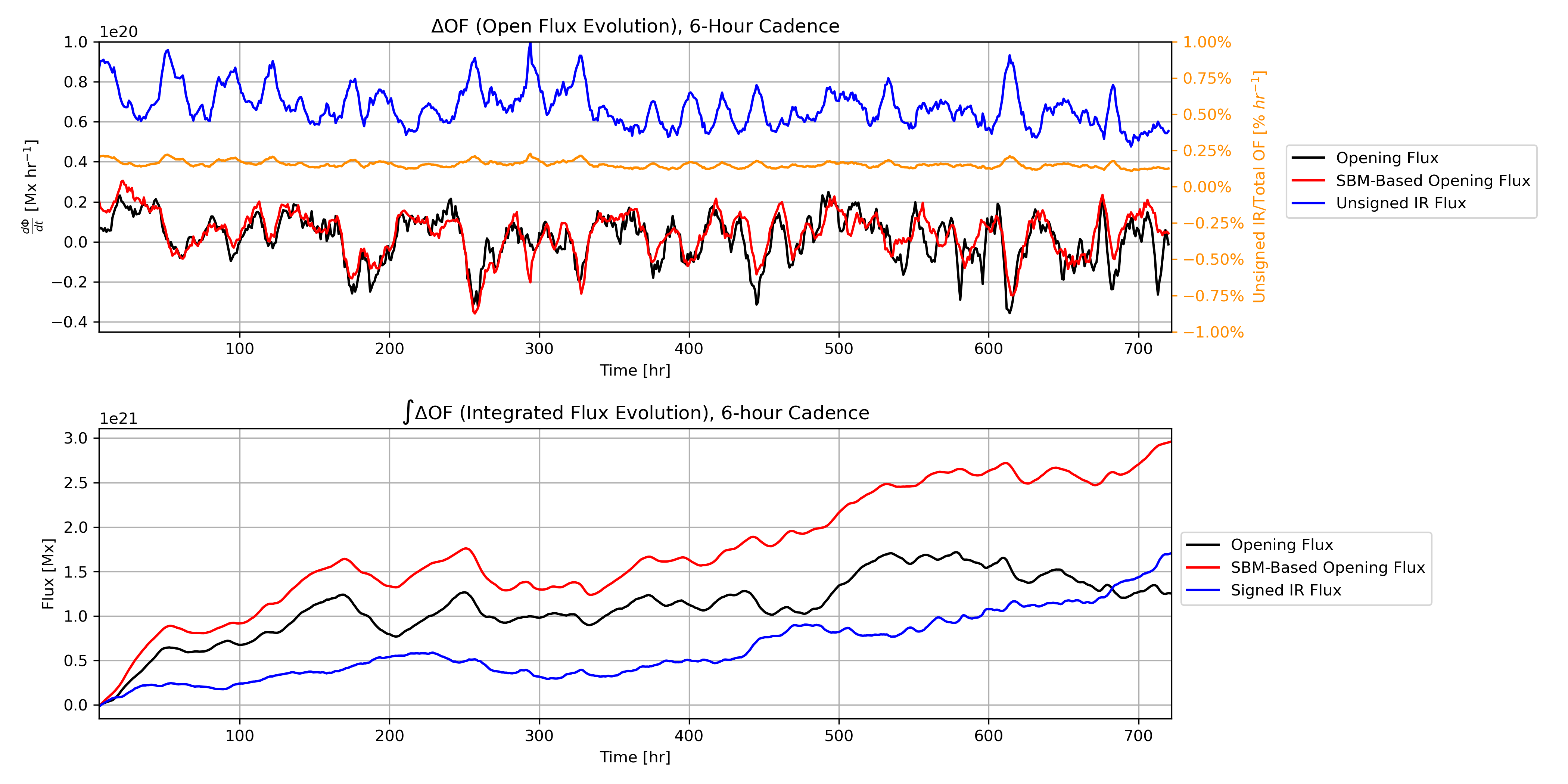}
    \caption{Top: plot showing the hourly change in open flux evolution for the three categories of SBM events (red), as well as the standard open flux calculation for the simulation (black). The unsigned difference between these two quantities is plotted as the {unsigned} interchange reconnection in blue. {The orange line corresponds to the ratio of the hourly unsigned IR to the total open flux, with associated units of ratio percent per hour.} Bottom: cumulative sums of flux evolution, both the total open flux and the flux associated with interchange reconnection.}
    \label{fig:of_calcs}
\end{figure}

The top plot of Figure~\ref{fig:of_calcs} shows the evolution of the change in open flux ($\Delta OF$) over the course of the simulation. The black line shows the standard $\Delta OF$ calculation for the simulation {-- that is, the differencing of 6-hour windows of the open field calculated from the model outputs}. The SBM-based $\Delta OF$ is plotted in red, and the \textit{unsigned} $\Delta OF$ associated with interchange reconnection is plotted in blue. Explanations of both derivations follow.

While the SBM calculation itself uses 6 hourly time steps to compute each map, the maps advance by a single time step for each calculation (i.e., one slip-back map covers hours 1-6, the next covers hours 2-7, etc.). Thus, the time derivative itself is still per hour. The $\Delta OF$ using the slip-back maps is computed using the following definition:
\begin{equation}
    \Delta OF_{SBM} = \frac{1}{2}[(FO_{dual}-FC_{dual})+(FO_{primary}-FC_{primary})],
\end{equation}where $\Delta OF_{SBM}$ is the change in the total open flux, $FO_{dual}$ is the flux measured as opening in the dual mapping, $FC_{dual}$ is the flux measured as closing in the dual mapping, $FO_{primary}$ is the flux measured as opening in the primary mapping, $FC_{primary}$ is the flux measured as closing in the primary mapping. The factor of one-half is included to remove the double-counting inherent in the categorization method.

One of the main goals of the SBM method is to trace regions where interchange reconnection is likely to occur, but this is challenging for several reasons. As discussed in Section \ref{sec:sbm}, the categorization system used here does not include interchange reconnection separately, largely due to the challenge posed by identifying it. The very nature of interchange reconnection complicates flux calculations, as there is no net change in the open flux. Additionally, resistive MHD is unable to realistically capture the microphysics involved in magnetic reconnection, but general dynamics can be well reproduced. The introduction of the slip surface also excludes detection of any closed field that lies below the slip surface's location, although it is of central importance. Flux emergence would cause failures in the calculation if the inner and outer boundaries were the surfaces from which the SBM was generated; the slip surfaces are located at heights corresponding to a characteristic radial advection distance (this advection speed is much lower for the inner boundary, which is why the inner slip surface is located only 0.01 $R_\odot$ above the inner boundary, which the outer slip surface is located 1 $R_\odot$ below the outer boundary). Therefore, there are many cases -- particularly involving small parasitic bipoles in coronal holes -- where very short field lines are excluded from the calculation entirely. However, interchange reconnection can be recovered using the values plotted in Figure~\ref{fig:of_calcs}. Using
\begin{equation}
    IR{_{US}} = \frac{1}{2}[FO_{dual}+FC_{dual}+FO_{primary}+FC_{primary}]-OF_{total},
\end{equation}where $OF$ is the open flux from the simulation, we recover the total {\textit{unsigned}} flux involved in interchange reconnection. {The \textit{signed} interchange reconnection flux can be found similarly:
\begin{equation}
    IR_{S} = \frac{1}{2}[(FO_{dual}-FC_{dual})+(FO_{primary}-FC_{primary})]-OF_{total},
\end{equation} which is equivalent to $\Delta OF_{SBM}-OF_{total}$.}
The bottom plot of Fig~\ref{fig:of_calcs} shows the cumulative open fluxes, and the cumulative {signed interchange reconnection flux is plotted in blue}.

\begin{figure}[htbp]
    \centering
    \includegraphics[width=0.85\linewidth]{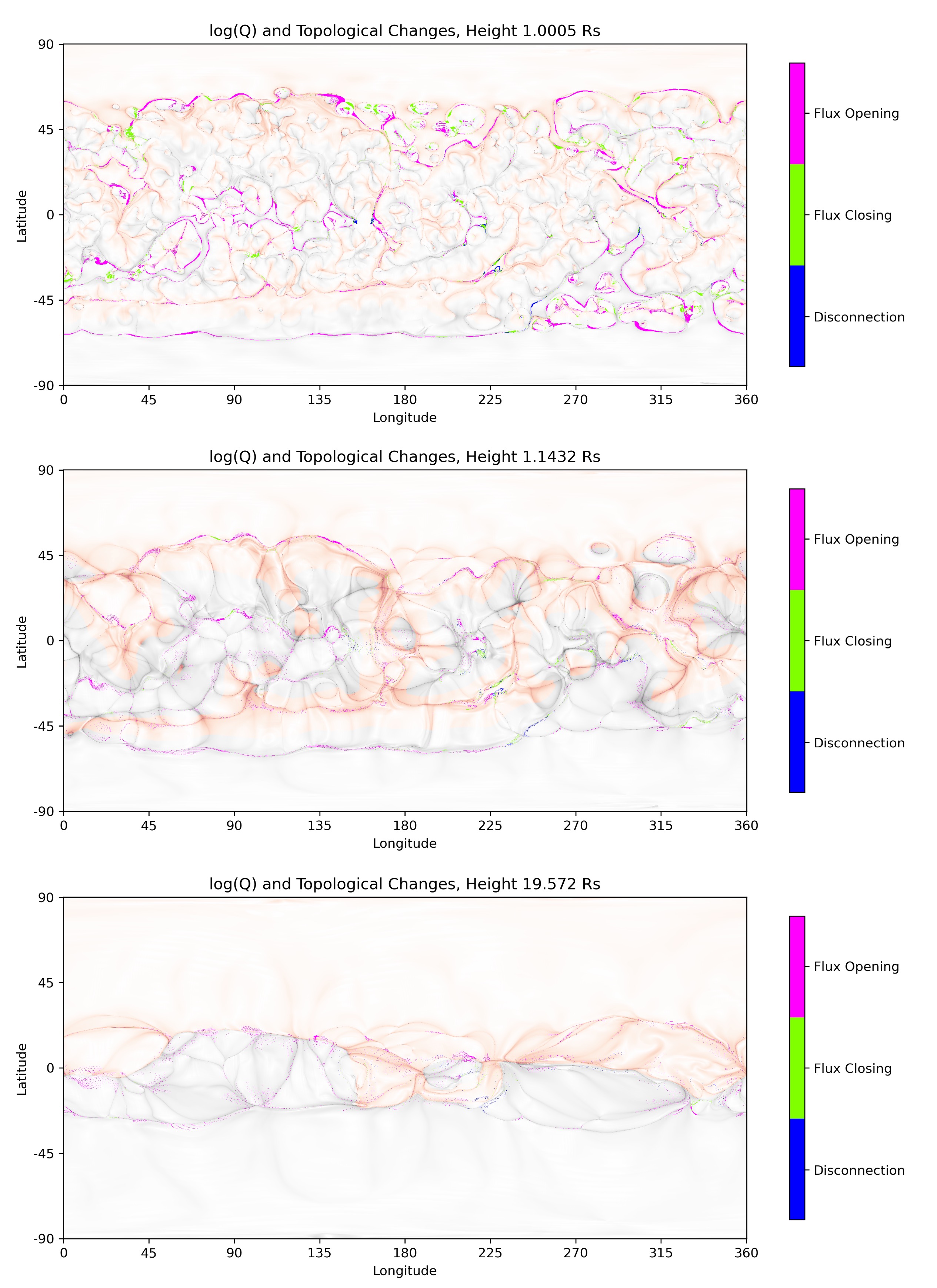}
    \caption{Overlay of crossing points of every field line involved in slip-back mapping associated with flux opening, closing, or disconnecting. The bottom layer of the image is a signed $log(Q)$ mapping of the magnetic field at the height listed in the title of the panel. These images show how the field lines that undergo topological change are entirely isolated to high-Q regions, as predicted by theory. An animation of this figure, showing the progression in increasing $r$ slices, is available online.}
    \label{fig:slogq_sbm}
\end{figure}

Figure~\ref{fig:slogq_sbm} shows three panels at the same time step, but at different heights within the simulation. The red/gray map shows the squashing factor (or Q), scaled logarithmically and signed for magnetic polarity. Over this is plotted the intersection with the plotted height of every field line involved in a SBM code, colored by category. The three heights shown here are extracted from the full animation (available in the online version of the paper) which shows the progression in height from 1 to 19 $R_\odot$ at a single time. This visualization provides perspective on how these field lines align with the major magnetic structures in the corona. As can be seen in the figure and the animation, the SBM field lines lie almost entirely along high-Q lines. While the over-plotted points appear widely distributed in the low corona, as the height increases it becomes obvious that the FO codes predominate along the heliospheric current sheet (seen as the main boundary between the red and gray sectors in the map), while the FC and {DC} codes tend to show up near quasi-separatrix layers (QSLs), or pseudostreamer arcs (visible as high-Q lines that lie entirely within one polarity). In theory, over time all three codes should occur along both types of arc. However, at a given time, it is more likely to observe this distribution. FO is often associated with the longest helmet streamer field lines growing beyond the slip surface, where their coding changes from closed to open. This is more common and less impulsive of an occurrence than helmet streamer pinch-off, where FC and DC codes would be expected to predominate locally. The high rates of interchange reconnection associated with the null-point topologies that plot as QSLs, on the other hand, are very likely to show a mix of all three bins at any given time, as seen here.

\section{Discussion}\label{sec:disc}

\subsection{Time-Dependent Analysis}
We have described the algorithm behind the SBM method and applied it to a fully time-dependent thermodynamic MHD simulation covering a month of solar evolution. The time dependence of the SBM method is central to its nature, and a necessary advance in MHD analysis tools. As our modeling capabilities continue to improve, so must our tools evolve to take the best advantage of the enhanced results. \textit{SBM is an inherently time-dependent metric, developed for time-dependent simulations.} As we show in Fig.~\ref{fig:of_calcs}, SBM tracks the open flux evolution in a time-dependent model without necessitating the assumption of a potential field, and it can also be used to find the OF at any one time. We have focused here on the full codes, since the investigation's goal was to trace the major trends in topological change, but more traditional quantities can also be recovered from a single mapping.

The method really excels in the aggregate, where binning multiple SBM codes together allows for the creation of maps and animations that show how and where flux is being processed throughout the corona and into the solar wind. The binning choices made in this paper are far from the only way to categorize these codes. Other categorizations would allow researchers to zero in on specific types of topological changes -- for instance, to focus on reconnection dynamics at the heliospheric current sheet like streamer blow-out or pinch-off events. Alternatively, one could focus solely on the codes that associate with small-scale interchange reconnection, to investigate the evolution of very small null-point topologies. Yet another avenue of research would be to map from the upper slip surface to gain insights on the open and disconnected field in the solar wind as it propagates outward. Some of these will be the focus of future work, but we include a wide range of ideas as inspiration to others in the adoption of this method for other time-dependent models.

\subsection{Solar Wind Implications}
While developing this methodology, we experimented with a broad range of cadences for the mappings, ranging from 3 minutes to 24 hours. We consistently found that -- while specific codes may shift with cadence -- categorized trends did not. For example, a high-Q arc that was predominantly populated with FO codes tended to remain an arc of predominantly FO codes, although more FO codes generally accumulated in the area as the cadence period increased. We chose the 24-hr cadence for Fig.~\ref{fig:slogq_sbm} since it allowed reasonable visibility of the patterns created by the mapping, while the rest of the plots illustrate slip-back calculations made with a 6-hr cadence. The 6-hr cadence was chosen as a time frame that agrees well with typical solar wind propagation speeds from the inner to the outer boundary of the calculation, allowing us to investigate the dynamics in the inner corona that contribute to the solar wind.

A coherent and compelling picture emerges from a comparison of the flux processed by the three main categories of topological change and the plot of the location of these changes with respect to a Q map. The total open flux for the simulation is around $4\cdot10^{22}$ Mx on average. This means that for any given {24-hour interval}, \textit{approximately {3.5}\% of the Sun's open flux is processed through interchange reconnection}, as defined by this investigation and using the values plotted in Fig.~\ref{fig:of_calcs}. It should be noted that this is specifically for a simulation scaled to approximate the conditions near solar minimum. During periods closer to solar maximum, greater magnetic flux emergence and more closely located regions of opposing polarities should lead to significantly higher proportions of open flux which has undergone interchange reconnection.

The field lines associated with SBM codes are widely distributed along high-Q lines low in the corona as shown in Fig.~\ref{fig:slogq_sbm}, with many appearing along QSLs. QSLs can be thought of as the projection of a null-point on the S-web, in much the same way as a null-point is itself the projection of a neutral line from the photosphere into the low corona. These arcs in the S-web indicate pseudo-streamers and smaller null-point topologies, and therefore are predicted to be strongly correlated to interchange reconnection. It is critical to note that Fig.~\ref{fig:of_calcs} also shows that significantly more flux is processed by interchange reconnection than by field-opening or field-closing methods. We see this correlation reflected in Fig.~\ref{fig:slogq_sbm}, where the field lines associated with topological change definitively clustered along the lanes of S-web arcs. 

Although this study used a 6-hr cadence, a shorter cadence would highlight any ``bursty'' reconnection driven by higher-frequency changes in the flows, and could be used to compare to the temporal signatures of related observed solar wind phenomena, such as periodic density structures \citep{Viall2015,DiMatteo2019,Kepko2024}. It should also be noted that this cadence does not separate individual reconnection events; it is reasonable to assume that many interchange reconnection events occur at these locations during this time, so the 1\% result should be considered a lower bound.

\subsection{Open Flux Implications}
SBM provides a more nuanced view of open flux evolution, compared to a traditional open flux map derived from cell-by-cell connectivity traces at a single time step. This is because it retains the identity of the flux regions as they move through the advection tracing, rather than simply comparing the static locations based upon the grid of the simulation. By preserving the flux element's identity, it simplifies certain questions, such as whether a region along a coronal hole boundary has merely moved or undergone a topological change. This is critical for improving the physical interpretation of simulation results, and can be used to better estimate reconnection rates, measure open flux evolution, and disentangle convolved mechanisms (e.g., rigid rotation of coronal holes).

While the results from the top panel in Fig.~\ref{fig:of_calcs} show that the interchange reconnection-related flux is a few percent of the total open flux {over any 24 hour interval}, the bottom panel also implies that \textit{this is a relatively high proportion of the total open flux change over time.} As mentioned previously, these results are derived from a simulation with relatively quiet solar conditions; a simulation such as the 2024 total solar eclipse prediction presented in \cite{Downs2025} is anticipated to produce significantly higher proportions of such wind, and this will be investigated in future work.

\subsection{Summary}

Here, we presented results of the slip-back mapping method over a Carrington rotation of time, with individual codes highlighting localized magnetic field dynamics, as well as aggregated categorized maps to illustrate overall trends. 

The single code cases allow isolation of interactions at the scale of single magnetic field lines and can help provide insights into changes at the smallest spatial and temporal scales of the simulation. While care should be taken when interpreting individual codes due to the constraints of the simulation's resolution, they can also allow for more rigorous visualization of the behavior of small structures like jets.

The aggregated maps show how coronal plasma from closed field is released as solar wind concentrated primarily along S-web arcs. The ability to derive interchange reconnection rates using our method is new, and provides insight into what proportion of the solar wind should be anticipated to show closed-field abundances. Unsigned interchange reconnection flux dominates over open flux changes at every time in the simulation, providing additional evidence that this is a dominant dynamic in the corona. We found that approximately {3.5}\% of the open flux is processed through interchange reconnection {over any 24 hour interval}, which corresponds to a relatively high proportion of the total open flux changes over time in the heliosphere. These results are based on a solar minimum-like simulation {with a minimal amount of free-energy injection}, so these proportions are expected to increase for a more active scenario.

This paper illustrates the utility and importance of using time-dependent analysis methods for a time-dependent 3-dimensional magnetohydrostatic simulation. The slip-back mapping method scales well for dynamics at various ranges, and the aggregation plotting presented here reduces some of the dimensionality of the method for easier visual parsing. This is one of many tools which should be developed to enhance our understanding of time-dependent simulations.

\begin{acknowledgments}
 We thank the NASA High-End Computing (HEC) Program through the NASA Advanced Supercomputing Division (NAS) at Ames Research Center for allocations on the Pleiades, Electra, and Aitken supercomputers and the NSF ACCESS program for allocations on the Expanse supercomputer at the San Diego Supercomputer Center (SDSC), which were used to run the simulations. This work was supported by the NASA Living With a Star Strategic Capabilities program (80NSSC22K0893), NSF SHINE program (AGS 2501333),  NASA Living With a Star Science program (80NSSC20K0192 and 80NSSC22K1021), and NSF PREEVENTS program (ICER 1854790). 
\end{acknowledgments}

\bibliography{sbm_bib}{}

@article{Titov2009,
  title = {Slip-squashing Factors as a Measure of Three-dimensional Magnetic Reconnection},
  volume = {693},
  ISSN = {1538-4357},
  url = {http://dx.doi.org/10.1088/0004-637X/693/1/1029},
  DOI = {10.1088/0004-637x/693/1/1029},
  number = {1},
  journal = {The Astrophysical Journal},
  publisher = {American Astronomical Society},
  author = {Titov,  V. S. and Forbes,  T. G. and Priest,  E. R. and Mikić,  Z. and Linker,  J. A.},
  year = {2009},
  month = mar,
  pages = {1029–1044}
}

@article{Lionello2020,
  title = {Slip-back Mapping as a Tracker of Topological Changes in Evolving Magnetic Configurations},
  volume = {891},
  ISSN = {1538-4357},
  url = {http://dx.doi.org/10.3847/1538-4357/ab68d9},
  DOI = {10.3847/1538-4357/ab68d9},
  number = {1},
  journal = {The Astrophysical Journal},
  publisher = {American Astronomical Society},
  author = {Lionello,  R. and Titov,  V. S. and Mikić,  Z. and Linker,  J. A.},
  year = {2020},
  month = feb,
  pages = {14}
}

@article{Downs2025,
  title = {A near-real-time data-assimilative model of the solar corona},
  volume = {388},
  ISSN = {1095-9203},
  url = {http://dx.doi.org/10.1126/science.adq0872},
  DOI = {10.1126/science.adq0872},
  number = {6753},
  journal = {Science},
  publisher = {American Association for the Advancement of Science (AAAS)},
  author = {Downs,  Cooper and Linker,  Jon A. and Caplan,  Ronald M. and Mason,  Emily I. and Riley,  Pete and Davidson,  Ryder and Reyes,  Andres and Palmerio,  Erika and Lionello,  Roberto and Turtle,  James and Ben-Nun,  Michal and Stulajter,  Miko M. and Titov,  Viacheslav S. and T\"{o}r\"{o}k,  Tibor and Upton,  Lisa A. and Attie,  Raphael and Jha,  Bibhuti K. and Arge,  Charles N. and Henney,  Carl J. and Valori,  Gherardo and Strecker,  Hanna and Calchetti,  Daniele and Germerott,  Dietmar and Hirzberger,  Johann and Suárez,  David Orozco and Rodríguez,  Julian Blanco and Solanki,  Sami K. and Cheng,  Xin and Wu,  Sizhe},
  year = {2025},
  month = jun,
  pages = {1306–1310}
}

@article{Mikic1999,
    title = {{Magnetohydrodynamic modeling of the global solar corona}},
    year = {1999},
    journal = {Physics of Plasmas},
    author = {Miki{\'{c}}, Zoran and Linker, Jon A. and Schnack, Dalton D. and Lionello, Roberto and Tarditi, Alfonso},
    number = {5},
    month = {5},
    pages = {2217--2224},
    volume = {6},
    url = {http://aip.scitation.org/doi/10.1063/1.873474},
    doi = {10.1063/1.873474},
    issn = {1070-664X}
}

@article{Lionello2013b,
    title = {{Magnetohydrodynamic simulations of interplanetary coronal mass ejections}},
    year = {2013},
    journal = {The Astrophysical Journal},
    author = {Lionello, Roberto and Downs, Cooper and Linker, Jon A. and T{\"{o}}r{\"{o}}k, Tibor and Riley, Pete and Miki{\'{c}}, Zoran},
    number = {1},
    month = {10},
    pages = {76},
    volume = {777},
    url = {https://iopscience.iop.org/article/10.1088/0004-637X/777/1/76},
    doi = {10.1088/0004-637X/777/1/76},
    issn = {0004-637X}
}

@article{Lionello2009,
    title = {{Multispectral emission of the sun during the first whole sun month: magnetohydrodynamic simulations}},
    year = {2009},
    journal = {The Astrophysical Journal},
    author = {Lionello, Roberto and Linker, Jon A. and Miki{\'{c}}, Zoran},
    number = {1},
    month = {1},
    pages = {902--912},
    volume = {690},
    url = {https://iopscience.iop.org/article/10.1088/0004-637X/690/1/902},
    doi = {10.1088/0004-637X/690/1/902},
    issn = {0004-637X}
}

@article{Riley2019,
    title = {{Predicting the Structure of the Solar Corona and Inner Heliosphere during Parker Solar Probe 's First Perihelion Pass}},
    year = {2019},
    journal = {The Astrophysical Journal},
    author = {Riley, Pete and Downs, Cooper and Linker, Jon A. and Mikic, Zoran and Lionello, Roberto and Caplan, Ronald M.},
    number = {2},
    month = {3},
    pages = {L15},
    volume = {874},
    url = {https://iopscience.iop.org/article/10.3847/2041-8213/ab0ec3},
    doi = {10.3847/2041-8213/ab0ec3},
    issn = {2041-8213}
}

@article{Downs2013,
    title = {{Probing the Solar Magnetic Field with a Sun-Grazing Comet}},
    year = {2013},
    journal = {Science},
    author = {Downs, Cooper and Linker, Jon A. and Miki{\'{c}}, Zoran and Riley, Pete and Schrijver, Carolus J. and Saint-Hilaire, Pascal},
    number = {6137},
    month = {6},
    pages = {1196--1199},
    volume = {340},
    url = {https://www.science.org/doi/10.1126/science.1236550},
    doi = {10.1126/science.1236550},
    issn = {0036-8075}
}

@article{Torok2018,
    title = {{Sun-to-Earth MHD Simulation of the 2000 July 14 “Bastille Day” Eruption}},
    year = {2018},
    journal = {The Astrophysical Journal},
    author = {T{\"{o}}r{\"{o}}k, Tibor and Downs, Cooper and Linker, Jon A. and Lionello, R. and Titov, Viacheslav S. and Miki{\'{c}}, Zoran and Riley, Pete and Caplan, Ronald M. and Wijaya, Janvier},
    number = {1},
    month = {3},
    pages = {75},
    volume = {856},
    url = {https://iopscience.iop.org/article/10.3847/1538-4357/aab36d},
    doi = {10.3847/1538-4357/aab36d},
    issn = {0004-637X}
}

@article{Riley2011,
    title = {{Global MHD Modeling of the Solar Corona and Inner Heliosphere for the Whole Heliosphere Interval}},
    year = {2011},
    journal = {Solar Physics},
    author = {Riley, P. and Lionello, R. and Linker, J. A. and Mikic, Z. and Luhmann, J. and Wijaya, J.},
    number = {1-2},
    month = {12},
    pages = {361--377},
    volume = {274},
    url = {http://link.springer.com/10.1007/s11207-010-9698-x},
    doi = {10.1007/s11207-010-9698-x},
    issn = {0038-0938}
}

@article{Downs2016,
    title = {{Closed-Field Coronal Heating Driven By Wave Turbulence}},
    year = {2016},
    journal = {The Astrophysical Journal},
    author = {Downs, Cooper and Lionello, Roberto and Miki{\'{c}}, Zoran and Linker, Jon A and Velli, Marco},
    number = {2},
    pages = {180},
    volume = {832},
    doi = {10.3847/0004-637x/832/2/180},
    issn = {15384357}
}

@article{Mikic2018PredictingEclipse,
    title = {{Predicting the corona for the 21 August 2017 total solar eclipse}},
    year = {2018},
    journal = {Nature Astronomy},
    author = {Miki{\'{c}}, Zoran and Downs, Cooper and Linker, Jon A. and Caplan, Ronald M. and Mackay, Duncan H. and Upton, Lisa A. and Riley, Pete and Lionello, Roberto and T{\"{o}}r{\"{o}}k, Tibor and Titov, Viacheslav S. and Wijaya, Janvier and Druckm{\"{u}}ller, Miloslav and Pasachoff, Jay M. and Carlos, Wendy},
    number = {11},
    month = {8},
    pages = {913--921},
    volume = {2},
    url = {https://www.nature.com/articles/s41550-018-0562-5},
    doi = {10.1038/s41550-018-0562-5},
    issn = {2397-3366}
}

@ARTICLE{telloni22,
       author = {{Telloni}, Daniele and {Zank}, Gary P. and {Stangalini}, Marco and {Downs}, Cooper and {Liang}, Haoming and {Nakanotani}, Masaru and {Andretta}, Vincenzo and {Antonucci}, Ester and {Sorriso-Valvo}, Luca and {Adhikari}, Laxman and {Zhao}, Lingling and {Marino}, Raffaele and {Susino}, Roberto and {Grimani}, Catia and {Fabi}, Michele and {D'Amicis}, Raffaella and {Perrone}, Denise and {Bruno}, Roberto and {Carbone}, Francesco and {Mancuso}, Salvatore and {Romoli}, Marco and {Deppo}, Vania Da and {Fineschi}, Silvano and {Heinzel}, Petr and {Moses}, John D. and {Naletto}, Giampiero and {Nicolini}, Gianalfredo and {Spadaro}, Daniele and {Teriaca}, Luca and {Frassati}, Federica and {Jerse}, Giovanna and {Landini}, Federico and {Pancrazzi}, Maurizio and {Russano}, Giuliana and {Sasso}, Clementina and {Biondo}, Ruggero and {Burtovoi}, Aleksandr and {Capuano}, Giuseppe E. and {Casini}, Chiara and {Casti}, Marta and {Chioetto}, Paolo and {De Leo}, Yara and {Giarrusso}, Marina and {Liberatore}, Alessandro and {Berghmans}, David and {Auch{\`e}re}, Fr{\'e}d{\'e}ric and {Cuadrado}, Regina Aznar and {Chitta}, Lakshmi P. and {Harra}, Louise and {Kraaikamp}, Emil and {Long}, David M. and {Mandal}, Sudip and {Parenti}, Susanna and {Pelouze}, Gabriel and {Peter}, Hardi and {Rodriguez}, Luciano and {Sch{\"u}hle}, Udo and {Schwanitz}, Conrad and {Smith}, Phil J. and {Verbeeck}, Cis and {Zhukov}, Andrei N.},
        title = "{Observation of a Magnetic Switchback in the Solar Corona}",
      journal = {The Astrophysical Journal},
         year = 2022,
        month = sep,
       volume = {936},
       number = {2},
          eid = {L25},
        pages = {L25},
          doi = {10.3847/2041-8213/ac8104},
archivePrefix = {arXiv},
       eprint = {2206.03090},
 primaryClass = {astro-ph.SR},
       adsurl = {https://ui.adsabs.harvard.edu/abs/2022ApJ...936L..25T}
}

@ARTICLE{antonucci23,
       author = {{Antonucci}, E. and {Downs}, C. and {Capuano}, G.~E. and {Spadaro}, D. and {Susino}, R. and {Telloni}, D. and {Andretta}, V. and {Da Deppo}, V. and {De Leo}, Y. and {Fineschi}, S. and {Frassetto}, F. and {Landini}, F. and {Naletto}, G. and {Nicolini}, G. and {Pancrazzi}, M. and {Romoli}, M. and {Stangalini}, M. and {Teriaca}, L. and {Uslenghi}, M.},
        title = "{Slow wind belt in the quiet solar corona}",
      journal = {Physics of Plasmas},
         year = 2023,
        month = feb,
       volume = {30},
       number = {2},
          eid = {022905},
        pages = {022905},
          doi = {10.1063/5.0132824},
archivePrefix = {arXiv},
       eprint = {2302.08385},
 primaryClass = {astro-ph.SR},
       adsurl = {https://ui.adsabs.harvard.edu/abs/2023PhPl...30b2905A}
}

@ARTICLE{downs2021,
       author = {{Downs}, Cooper and {Warmuth}, Alexander and {Long}, David M. and {Bloomfield}, D. Shaun and {Kwon}, Ryun-Young and {Veronig}, Astrid M. and {Vourlidas}, Angelos and {Vr{\v{s}}nak}, Bojan},
        title = "{Validation of Global EUV Wave MHD Simulations and Observational Techniques}",
      journal = {The Astrophysical Journal},
         year = 2021,
        month = apr,
       volume = {911},
       number = {2},
          eid = {118},
        pages = {118},
          doi = {10.3847/1538-4357/abea78},
       adsurl = {https://ui.adsabs.harvard.edu/abs/2021ApJ...911..118D}
}

@article{Yeates2018,
    title = {{Global Non-Potential Magnetic Models of the Solar Corona During the March 2015 Eclipse}},
    year = {2018},
    journal = {Space Science Reviews},
    author = {Yeates, Anthony R. and Amari, Tahar and Contopoulos, Ioannis and Feng, Xueshang and Mackay, Duncan H. and Miki{\'{c}}, Zoran and Wiegelmann, Thomas and Hutton, Joseph and Lowder, Christopher A. and Morgan, Huw and Petrie, Gordon and Rachmeler, Laurel A. and Upton, Lisa A. and Canou, Aurelien and Chopin, Pierre and Downs, Cooper and Druckm{\"{u}}ller, Miloslav and Linker, Jon A. and Seaton, Daniel B. and T{\"{o}}r{\"{o}}k, Tibor},
    number = {5},
    month = {8},
    pages = {99},
    volume = {214},
    url = {http://link.springer.com/10.1007/s11214-018-0534-1},
    doi = {10.1007/s11214-018-0534-1},
    issn = {0038-6308}
}

@article{Lionello2023GlobalCorona,
       author = {{Lionello}, Roberto and {Downs}, Cooper and {Mason}, Emily I. and {Linker}, Jon A. and {Caplan}, Ronald M. and {Riley}, Pete and {Titov}, Viacheslav S. and {DeRosa}, Marc L.},
        title = "{Global MHD Simulations of the Time-dependent Corona}",
      journal = {The Astrophysical Journal},
         year = 2023,
        month = dec,
       volume = {959},
       number = {2},
          eid = {77},
        pages = {77},
          doi = {10.3847/1538-4357/ad00be},
archivePrefix = {arXiv},
       eprint = {2306.12551},
 primaryClass = {astro-ph.SR},
       adsurl = {https://ui.adsabs.harvard.edu/abs/2023ApJ...959...77L}
}

@ARTICLE{Mason2023TDC,
       author = {{Mason}, Emily I. and {Lionello}, Roberto and {Downs}, Cooper and {Linker}, Jon A. and {Caplan}, Ronald M. and {DeRosa}, Marc L.},
        title = "{Time-dependent Dynamics of the Corona}",
      journal = {The Astrophysical Journal Letters},
         year = 2023,
        month = dec,
       volume = {959},
       number = {1},
          eid = {L4},
        pages = {L4},
          doi = {10.3847/2041-8213/ad00bd},
archivePrefix = {arXiv},
       eprint = {2306.11956},
 primaryClass = {astro-ph.SR},
       adsurl = {https://ui.adsabs.harvard.edu/abs/2023ApJ...959L...4M}
}

@article{schrijver03b,
    title = {{Photospheric and heliospheric magnetic fields}},
    year = {2003},
    journal = {Solar Physics},
    author = {Schrijver, C.J. and DeRosa, M.L.},
    month = {1},
    pages = {165--200},
    volume = {212},
    doi = {10.1023/A:1022908504100}
}

@article{Linker2017,
    title = {{The Open Flux Problem}},
    year = {2017},
    journal = {The Astrophysical Journal},
    author = {Linker, J. A. and Caplan, R. M. and Downs, C. and Riley, P. and Mikic, Z. and Lionello, R. and Henney, C. J. and Arge, C. N. and Liu, Y. and Derosa, M. L. and Yeates, A. and Owens, M. J.},
    number = {1},
    pages = {70},
    volume = {848},
    publisher = {IOP Publishing},
    url = {http://dx.doi.org/10.3847/1538-4357/aa8a70},
    doi = {10.3847/1538-4357/aa8a70},
    issn = {15384357},
    arxivId = {1708.02342}
}

@ARTICLE{Lowder2017,
       author = {{Lowder}, Chris and {Qiu}, Jiong and {Leamon}, Robert},
        title = "{Coronal Holes and Open Magnetic Flux over Cycles 23 and 24}",
      journal = {\solphys},
         year = 2017,
        month = jan,
       volume = {292},
       number = {1},
          eid = {18},
        pages = {18},
          doi = {10.1007/s11207-016-1041-8},
archivePrefix = {arXiv},
       eprint = {1612.07595},
 primaryClass = {astro-ph.SR},
       adsurl = {https://ui.adsabs.harvard.edu/abs/2017SoPh..292...18L}
}

@ARTICLE{Wallace2019,
       author = {{Wallace}, S. and {Arge}, C.~N. and {Pattichis}, M. and {Hock-Mysliwiec}, R.~A. and {Henney}, C.~J.},
        title = "{Estimating Total Open Heliospheric Magnetic Flux}",
      journal = {\solphys},
         year = 2019,
        month = feb,
       volume = {294},
       number = {2},
          eid = {19},
        pages = {19},
          doi = {10.1007/s11207-019-1402-1},
archivePrefix = {arXiv},
       eprint = {1903.12613},
 primaryClass = {astro-ph.SR},
       adsurl = {https://ui.adsabs.harvard.edu/abs/2019SoPh..294...19W}
}

@article{Owens2014,
  title = {Solar cycle evolution of dipolar and pseudostreamer belts and their relation to the slow solar wind},
  volume = {119},
  ISSN = {2169-9402},
  url = {http://dx.doi.org/10.1002/2013JA019412},
  DOI = {10.1002/2013ja019412},
  number = {1},
  journal = {Journal of Geophysical Research: Space Physics},
  publisher = {American Geophysical Union (AGU)},
  author = {Owens,  M. J. and Crooker,  N. U. and Lockwood,  M.},
  year = {2014},
  month = jan,
  pages = {36–46}
}

@ARTICLE{Viall2020,
       author = {{Viall}, Nicholeen M. and {Borovsky}, Joseph E.},
        title = "{Nine Outstanding Questions of Solar Wind Physics}",
      journal = {Journal of Geophysical Research (Space Physics)},
         year = 2020,
        month = jul,
       volume = {125},
       number = {7},
          eid = {e26005},
        pages = {e26005},
          doi = {10.1029/2018JA02600510.1002/essoar.10502606.1},
       adsurl = {https://ui.adsabs.harvard.edu/abs/2020JGRA..12526005V}
}

@ARTICLE{Bale2019,
       author = {{Bale}, S.~D. and {Badman}, S.~T. and {Bonnell}, J.~W. and {Bowen}, T.~A. and {Burgess}, D. and {Case}, A.~W. and {Cattell}, C.~A. and {Chandran}, B.~D.~G. and {Chaston}, C.~C. and {Chen}, C.~H.~K. and {Drake}, J.~F. and {de Wit}, T. Dudok and {Eastwood}, J.~P. and {Ergun}, R.~E. and {Farrell}, W.~M. and {Fong}, C. and {Goetz}, K. and {Goldstein}, M. and {Goodrich}, K.~A. and {Harvey}, P.~R. and {Horbury}, T.~S. and {Howes}, G.~G. and {Kasper}, J.~C. and {Kellogg}, P.~J. and {Klimchuk}, J.~A. and {Korreck}, K.~E. and {Krasnoselskikh}, V.~V. and {Krucker}, S. and {Laker}, R. and {Larson}, D.~E. and {MacDowall}, R.~J. and {Maksimovic}, M. and {Malaspina}, D.~M. and {Martinez-Oliveros}, J. and {McComas}, D.~J. and {Meyer-Vernet}, N. and {Moncuquet}, M. and {Mozer}, F.~S. and {Phan}, T.~D. and {Pulupa}, M. and {Raouafi}, N.~E. and {Salem}, C. and {Stansby}, D. and {Stevens}, M. and {Szabo}, A. and {Velli}, M. and {Woolley}, T. and {Wygant}, J.~R.},
        title = "{Highly structured slow solar wind emerging from an equatorial coronal hole}",
      journal = {\nat},
         year = 2019,
        month = dec,
       volume = {576},
       number = {7786},
        pages = {237-242},
          doi = {10.1038/s41586-019-1818-7},
       adsurl = {https://ui.adsabs.harvard.edu/abs/2019Natur.576..237B}
}

@ARTICLE{Borovsky2008,
       author = {{Borovsky}, Joseph E.},
        title = "{Flux tube texture of the solar wind: Strands of the magnetic carpet at 1 AU?}",
      journal = {Journal of Geophysical Research (Space Physics)},
         year = 2008,
        month = aug,
       volume = {113},
       number = {A8},
          eid = {A08110},
        pages = {A08110},
          doi = {10.1029/2007JA012684},
       adsurl = {https://ui.adsabs.harvard.edu/abs/2008JGRA..113.8110B}
}

@ARTICLE{Geiss1995,
       author = {{Geiss}, J. and {Gloeckler}, G. and {von Steiger}, R.},
        title = "{Origin of the Solar Wind From Composition Data}",
      journal = {\ssr},
         year = 1995,
        month = apr,
       volume = {72},
       number = {1-2},
        pages = {49-60},
          doi = {10.1007/BF00768753},
       adsurl = {https://ui.adsabs.harvard.edu/abs/1995SSRv...72...49G}
}

@ARTICLE{Antiochos2011,
       author = {{Antiochos}, S.~K. and {Miki{\'c}}, Z. and {Titov}, V.~S. and {Lionello}, R. and {Linker}, J.~A.},
        title = "{A Model for the Sources of the Slow Solar Wind}",
      journal = {\apj},
         year = 2011,
        month = apr,
       volume = {731},
       number = {2},
          eid = {112},
        pages = {112},
          doi = {10.1088/0004-637X/731/2/112},
archivePrefix = {arXiv},
       eprint = {1102.3704},
 primaryClass = {astro-ph.SR},
       adsurl = {https://ui.adsabs.harvard.edu/abs/2011ApJ...731..112A}
}

@ARTICLE{Wang1998,
       author = {{Wang}, Y. -M. and {Sheeley}, Jr., N.~R. and {Walters}, J.~H. and {Brueckner}, G.~E. and {Howard}, R.~A. and {Michels}, D.~J. and {Lamy}, P.~L. and {Schwenn}, R. and {Simnett}, G.~M.},
        title = "{Origin of Streamer Material in the Outer Corona}",
      journal = {\apjl},
         year = 1998,
        month = may,
       volume = {498},
       number = {2},
        pages = {L165-L168},
          doi = {10.1086/311321},
       adsurl = {https://ui.adsabs.harvard.edu/abs/1998ApJ...498L.165W}
}

@ARTICLE{Laming2015,
       author = {{Laming}, J. Martin},
        title = "{The FIP and Inverse FIP Effects in Solar and Stellar Coronae}",
      journal = {Living Reviews in Solar Physics},
         year = 2015,
        month = dec,
       volume = {12},
       number = {1},
          eid = {2},
        pages = {2},
          doi = {10.1007/lrsp-2015-2},
archivePrefix = {arXiv},
       eprint = {1504.08325},
 primaryClass = {astro-ph.SR},
       adsurl = {https://ui.adsabs.harvard.edu/abs/2015LRSP...12....2L}
}

@article{West2023,
author = {West, Matthew J. and Seaton, Daniel B. and Wexler, David B. and Raymond, John C. and {Del Zanna}, Giulio and Rivera, Yeimy J. and Kobelski, Adam R. and Chen, Bin and DeForest, Craig and Golub, Leon and Caspi, Amir and Gilly, Chris R. and Kooi, Jason E. and Meyer, Karen A. and Alterman, Benjamin L. and Alzate, Nathalia and Andretta, Vincenzo and Auch{\`{e}}re, Fr{\'{e}}d{\'{e}}ric and Banerjee, Dipankar and Berghmans, David and Chamberlin, Phillip and Chitta, Lakshmi Pradeep and Downs, Cooper and Giordano, Silvio and Harra, Louise and Higginson, Aleida and Howard, Russell A. and Kumar, Pankaj and Mason, Emily and Mason, James P. and Morton, Richard J. and Nykyri, Katariina and Patel, Ritesh and Rachmeler, Laurel and Reardon, Kevin P. and Reeves, Katharine K. and Savage, Sabrina and Thompson, Barbara J. and {Van Kooten}, Samuel J. and Viall, Nicholeen M. and Vourlidas, Angelos and Zhukov, Andrei N.},
doi = {10.1007/s11207-023-02170-1},
issn = {0038-0938},
journal = {Solar Physics},
month = {jun},
number = {6},
pages = {78},
title = {{Defining the Middle Corona}},
url = {https://link.springer.com/10.1007/s11207-023-02170-1},
volume = {298},
year = {2023}
}

@ARTICLE{Parker1965,
       author = {{Parker}, E.~N.},
        title = "{Dynamical Theory of the Solar Wind}",
      journal = {\ssr},
         year = 1965,
        month = sep,
       volume = {4},
       number = {5-6},
        pages = {666-708},
          doi = {10.1007/BF00216273},
       adsurl = {https://ui.adsabs.harvard.edu/abs/1965SSRv....4..666P}
}

@ARTICLE{Cranmer2005,
       author = {{Cranmer}, S.~R. and {van Ballegooijen}, A.~A.},
        title = "{On the Generation, Propagation, and Reflection of Alfv{\'e}n Waves from the Solar Photosphere to the Distant Heliosphere}",
      journal = {\apjs},
         year = 2005,
        month = feb,
       volume = {156},
       number = {2},
        pages = {265-293},
          doi = {10.1086/426507},
archivePrefix = {arXiv},
       eprint = {astro-ph/0410639},
 primaryClass = {astro-ph},
       adsurl = {https://ui.adsabs.harvard.edu/abs/2005ApJS..156..265C}
}

@ARTICLE{Gosling1999,
       author = {{Gosling}, J.~T. and {Pizzo}, V.~J.},
        title = "{Formation and Evolution of Corotating Interaction Regions and their Three Dimensional Structure}",
      journal = {\ssr},
         year = 1999,
        month = jul,
       volume = {89},
        pages = {21-52},
          doi = {10.1023/A:1005291711900},
       adsurl = {https://ui.adsabs.harvard.edu/abs/1999SSRv...89...21G}
}

@ARTICLE{Karimabadi2013,
       author = {{Karimabadi}, H. and {Roytershteyn}, V. and {Wan}, M. and {Matthaeus}, W.~H. and {Daughton}, W. and {Wu}, P. and {Shay}, M. and {Loring}, B. and {Borovsky}, J. and {Leonardis}, E. and {Chapman}, S.~C. and {Nakamura}, T.~K.~M.},
        title = "{Coherent structures, intermittent turbulence, and dissipation in high-temperature plasmas}",
      journal = {Physics of Plasmas},
         year = 2013,
        month = jan,
       volume = {20},
       number = {1},
          eid = {012303},
        pages = {012303},
          doi = {10.1063/1.4773205},
       adsurl = {https://ui.adsabs.harvard.edu/abs/2013PhPl...20a2303K}
}

@article{Titov2011,
  title = {MAGNETIC TOPOLOGY OF CORONAL HOLE LINKAGES},
  volume = {731},
  ISSN = {1538-4357},
  url = {http://dx.doi.org/10.1088/0004-637X/731/2/111},
  DOI = {10.1088/0004-637x/731/2/111},
  number = {2},
  journal = {The Astrophysical Journal},
  publisher = {American Astronomical Society},
  author = {Titov,  V. S. and Mikić,  Z. and Linker,  J. A. and Lionello,  R. and Antiochos,  S. K.},
  year = {2011},
  month = mar,
  pages = {111}
}

@ARTICLE{Antiochos2012,
       author = {{Antiochos}, Spiro K. and {Linker}, Jon A. and {Lionello}, Roberto and {Miki{\'c}}, Zoran and {Titov}, Viacheslav and {Zurbuchen}, Thomas H.},
        title = "{The Structure and Dynamics of the Corona{\textemdash}Heliosphere Connection}",
      journal = {\ssr},
         year = 2012,
        month = nov,
       volume = {172},
       number = {1-4},
        pages = {169-185},
          doi = {10.1007/s11214-011-9795-7},
       adsurl = {https://ui.adsabs.harvard.edu/abs/2012SSRv..172..169A}
}

@ARTICLE{Chitta2023,
       author = {{Chitta}, L.~P. and {Seaton}, D.~B. and {Downs}, C. and {DeForest}, C.~E. and {Higginson}, A.~K.},
        title = "{Direct observations of a complex coronal web driving highly structured slow solar wind}",
      journal = {Nature Astronomy},
         year = 2023,
        month = feb,
       volume = {7},
        pages = {133-141},
          doi = {10.1038/s41550-022-01834-5},
archivePrefix = {arXiv},
       eprint = {2211.13283},
 primaryClass = {astro-ph.SR},
       adsurl = {https://ui.adsabs.harvard.edu/abs/2023NatAs...7..133C}
}

@ARTICLE{Baker2023,
       author = {{Baker}, D. and {D{\'e}moulin}, P. and {Yardley}, S.~L. and {Mihailescu}, T. and {van Driel-Gesztelyi}, L. and {D'Amicis}, R. and {Long}, D.~M. and {To}, A.~S.~H. and {Owen}, C.~J. and {Horbury}, T.~S. and {Brooks}, D.~H. and {Perrone}, D. and {French}, R.~J. and {James}, A.~W. and {Janvier}, M. and {Matthews}, S. and {Stangalini}, M. and {Valori}, G. and {Smith}, P. and {Cuadrado}, R. Aznar and {Peter}, H. and {Schuehle}, U. and {Harra}, L. and {Barczynski}, K. and {Berghmans}, D. and {Zhukov}, A.~N. and {Rodriguez}, L. and {Verbeeck}, C.},
        title = "{Observational Evidence of S-web Source of the Slow Solar Wind}",
      journal = {\apj},
         year = 2023,
        month = jun,
       volume = {950},
       number = {1},
          eid = {65},
        pages = {65},
          doi = {10.3847/1538-4357/acc653},
archivePrefix = {arXiv},
       eprint = {2303.12192},
 primaryClass = {astro-ph.SR},
       adsurl = {https://ui.adsabs.harvard.edu/abs/2023ApJ...950...65B}
}

@ARTICLE{Wallace2025,
       author = {{Wallace}, Samantha and {Higginson}, Aleida K. and {Simpson}, David G. and {Viall}, Nicholeen M. and {Wyper}, Peter and {Arge}, C. Nick},
        title = "{Deriving the Coronal Separatrix-Web With the WSA Model}",
      journal = {Journal of Geophysical Research (Space Physics)},
         year = 2025,
        month = aug,
       volume = {130},
       number = {8},
          eid = {e2025JA033743},
        pages = {e2025JA033743},
          doi = {10.1029/2025JA03374310.22541/essoar.173724287.72753854/v1},
       adsurl = {https://ui.adsabs.harvard.edu/abs/2025JGRA..13033743W}
}

@ARTICLE{Caplan2025,
       author = {{Caplan}, Ronald M. and {Stulajter}, Miko M. and {Linker}, Jon A. and {Downs}, Cooper and {Upton}, Lisa A. and {Jha}, Bibhuti Kumar and {Attie}, Raphael and {Arge}, Charles N. and {Henney}, Carl J.},
        title = "{Open-source Flux Transport (OFT). I. HipFT{\textendash}High-performance Flux Transport}",
      journal = {\apjs},
         year = 2025,
        month = may,
       volume = {278},
       number = {1},
          eid = {24},
        pages = {24},
          doi = {10.3847/1538-4365/adc080},
archivePrefix = {arXiv},
       eprint = {2501.06377},
 primaryClass = {astro-ph.SR},
       adsurl = {https://ui.adsabs.harvard.edu/abs/2025ApJS..278...24C}
}

@ARTICLE{Upton2024,
       author = {{Upton}, Lisa A. and {Ugarte-Urra}, Ignacio and {Warren}, Harry P. and {Hathaway}, David H.},
        title = "{The Advective Flux Transport Model: Improving the Far Side with Active Regions Observed by STEREO 304 {\r{A}}}",
      journal = {\apj},
         year = 2024,
        month = jun,
       volume = {968},
       number = {2},
          eid = {114},
        pages = {114},
          doi = {10.3847/1538-4357/ad40a5},
archivePrefix = {arXiv},
       eprint = {2404.04280},
 primaryClass = {astro-ph.SR},
       adsurl = {https://ui.adsabs.harvard.edu/abs/2024ApJ...968..114U}
}

@ARTICLE{Viall2015,
       author = {{Viall}, Nicholeen M. and {Vourlidas}, Angelos},
        title = "{Periodic Density Structures and the Origin of the Slow Solar Wind}",
      journal = {\apj},
         year = 2015,
        month = jul,
       volume = {807},
       number = {2},
          eid = {176},
        pages = {176},
          doi = {10.1088/0004-637X/807/2/176},
       adsurl = {https://ui.adsabs.harvard.edu/abs/2015ApJ...807..176V}
}

@ARTICLE{DiMatteo2019,
       author = {{Di Matteo}, S. and {Viall}, N.~M. and {Kepko}, L. and {Wallace}, S. and {Arge}, C.~N. and {MacNeice}, P.},
        title = "{Helios Observations of Quasiperiodic Density Structures in the Slow Solar Wind at 0.3, 0.4, and 0.6 AU}",
      journal = {Journal of Geophysical Research (Space Physics)},
         year = 2019,
        month = feb,
       volume = {124},
       number = {2},
        pages = {837-860},
          doi = {10.1029/2018JA026182},
       adsurl = {https://ui.adsabs.harvard.edu/abs/2019JGRA..124..837D}
}

@ARTICLE{Kepko2024,
       author = {{Kepko}, L. and {Viall}, N.~M. and {DiMatteo}, S.},
        title = "{Periodic Mesoscale Density Structures Comprise a Significant Fraction of the Solar Wind and Are Formed at the Sun}",
      journal = {Journal of Geophysical Research (Space Physics)},
         year = 2024,
        month = jan,
       volume = {129},
       number = {1},
          eid = {e2023JA031403},
        pages = {e2023JA031403},
          doi = {10.1029/2023JA031403},
       adsurl = {https://ui.adsabs.harvard.edu/abs/2024JGRA..12931403K}
}

@article{Wyper2018,
author = {Wyper, P. F. and DeVore, C. R. and Antiochos, S. K.},
doi = {10.3847/1538-4357/aa9ffc},
issn = {1538-4357},
journal = {The Astrophysical Journal},
number = {2},
pages = {98},
publisher = {IOP Publishing},
title = {{A Breakout Model for Solar Coronal Jets with Filaments}},
url = {http://dx.doi.org/10.3847/1538-4357/aa9ffc},
volume = {852},
year = {2018}
}

@INPROCEEDINGS{argeetal2013,
       author = {{Arge}, C. Nick and {Henney}, Carl J. and {Hernandez}, Irene Gonzalez and
         {Toussaint}, W. Alex and {Koller}, Josef and {Godinez}, Humberto C.},
        title = "{Modeling the corona and solar wind using ADAPT maps that include far-side observations}",
    booktitle = {Solar Wind 13},
         year = 2013,
       editor = {{Zank}, Gary P. and {Borovsky}, Joe and {Bruno}, Roberto and
         {Cirtain}, Jonathan and {Cranmer}, Steve and {Elliott}, Heather and
         {Giacalone}, Joe and {Gonzalez}, Walter and {Li}, Gang and
         {Marsch}, Eckart and {Moebius}, Ebehard and {Pogorelov}, Nick and
         {Spann}, Jim and {Verkhoglyadova}, Olga},
       series = {American Institute of Physics Conference Series},
       volume = {1539},
        month = jun,
        pages = {11-14},
          doi = {10.1063/1.4810977},
       adsurl = {https://ui.adsabs.harvard.edu/abs/2013AIPC.1539...11A}
}

@article{linkeretal2003,
	Adsurl = {http://adsabs.harvard.edu/abs/2003PhPl...10.1971L},
	Author = {{Linker}, J.~A. and {Miki{\'c}}, Z. and {Lionello}, R. and {Riley}, P. and {Amari}, T. and {Odstrcil}, D.},
	Doi = {10.1063/1.1563668},
	Journal = {Phys. Plasmas},
	Month = may,
	Pages = {1971-1978},
	Title = {{Flux cancellation and coronal mass ejections}},
	Volume = 10,
	Year = 2003}

@ARTICLE{linkeretal2021,
       author = {{Linker}, Jon A. and {Heinemann}, Stephan G. and {Temmer}, Manuela and {Owens}, Mathew J. and {Caplan}, Ronald M. and {Arge}, Charles N. and {Asvestari}, Eleanna and {Delouille}, Veronique and {Downs}, Cooper and {Hofmeister}, Stefan J. and {Jebaraj}, Immanuel C. and {Madjarska}, Maria S. and {Pinto}, Rui F. and {Pomoell}, Jens and {Samara}, Evangelia and {Scolini}, Camilla and {Vr{\v{s}}nak}, Bojan},
        title = "{Coronal Hole Detection and Open Magnetic Flux}",
      journal = {\apj},
         year = 2021,
        month = sep,
       volume = {918},
       number = {1},
          eid = {21},
        pages = {21},
          doi = {10.3847/1538-4357/ac090a},
archivePrefix = {arXiv},
       eprint = {2103.05837},
 primaryClass = {astro-ph.SR},
       adsurl = {https://ui.adsabs.harvard.edu/abs/2021ApJ...918...21L}
}
\bibliographystyle{aasjournalv7}

\appendix

\section{Details of the Slip-back Mapping Algorithm}\label{appendix}
The algorithm described here is the natural evolution of that outlined in \cite{Titov2009} and expanded in \cite{Lionello2020}. Two slip surfaces $R_0$ and $R_1$ are defined close to the lower and upper radial boundary respectively ($R_\odot \lesssim R_0 \ll R_1< R_\mathrm{up}$), and at two times: $t_1$ (the present), and $t_0<t_1$ (the past). The algorithm involves multiple steps, alternating tracing magnetic field lines and advecting points. Magnetic field lines are traced from different points at $t_1$ or $t_0$ by solving
\begin{equation}
    \frac{d \mathbf{x}}{d s} = \frac{\mathbf{B}(\mathbf{x})}{B(\mathbf{x})},
\end{equation}
with the predictor-corrector method. 
Here $\mathbf{x}$ is the position in the computational domain, $s$ the distance along the
field line. The magnetic field $\mathbf{B}$ for the time in question
is linearly interpolated in space from the values
provided by the MHD simulation. 
The advection of points between $t_0$ and $t_1$ is calculated with a similar predictor-corrector method 
\begin{equation}
    \frac{d \mathbf{x}}{d t} = \mathbf{\tilde{v}}(\mathbf{x},t).
\end{equation}
The flow $\mathbf{\tilde{v}}$ must be such that the final point belongs to the same field line as the initial point. Since reconnection is expected to occur between  $R_0$ and $R_1$, it is 
desirable that $\mathbf{x}$ always be close to the slip surface from which it originates and far from the polarity inversion line. Therefore, the choice $\mathbf{\tilde{v}}=\mathbf{v}$ (i.e., same flow as in the MHD simulation) is a poor one, as flux emergence/submergence and the solar wind may take a point far away from $R_0$ and $R_1$, respectively. A better choice is $\mathbf{v}_\mathrm{FP}$ as described in Eq.~(2) of
\citet{Lionello2020}. However, $\mathbf{v}_\mathrm{TFL}$ of Eq.~(8) of \citet{Lionello2020} is still better, since it allows
for general emergence of submergence of fluxes at any boundary. In the present calculation, 
we have used a similar formulation to   $\mathbf{v}_\mathrm{TFL}$ obtained by first taking the flow perpendicular to the 
magnetic field, $\mathbf{v}_\perp$:
\begin{equation}
    \mathbf{\tilde{v}}= \mathbf{v}_\perp 
        - \frac{v_{\perp r}\, B_{r}\, \mathbf{B}}{{B_{r}^2}+\epsilon^2}
  .
\end{equation}
As in \citet{Lionello2020} a small $\epsilon$ factor is introduced to eliminate
the singularity at the polarity inversion line. $\mathbf{\tilde{v}}$ is evaluated by linearly interpolating the fields of the MHD simulation
in space and time. The $[t_0,t_1]$ time interval is divided into $N$ smaller $\Delta t$ sub-intervals, corresponding to the cadence of field dumps in the simulation. Thus, for a given
point $\mathbf{x}$, $t\in [t_0+n\Delta t,t_0+(n+1)\Delta t]$, and $0\leq n < N$, we have:
\begin{eqnarray}
\mathbf{\tilde{v}}(t)&=&(1-\alpha)\mathbf{\tilde{v}}(t_0+n\Delta t)+
\alpha \mathbf{\tilde{v}}(t_0+(n+1)\Delta t) \nonumber \\
\alpha& =& \frac{t-(t_0+n\Delta t)}{\Delta t}
\end{eqnarray}
As in \citet{Lionello2020}, we use five steps \citep[][used four]{Titov2009}. For the primary slip-back mapping (Fig.~\ref{fig:mapping}a), they are the following:
\begin{enumerate}
    \item At $t=t_1$, we trace a magnetic field line from the initial point P1 located on one
    of the slip surfaces and determine the connectivity: if the field line reaches the other
    slip surface, the connectivity is open (O). Otherwise, the connectivity
 is recorded as closed (C, if both footpoints of the field line are at $R_0$) or
    disconnected (D, if both footpoints are at $R_1$).
    \item The end-point of the field line is then advected back in time from $t_1$ to $t_0$ with the  $\mathbf{\tilde{v}}(t)$ flow. We label the final point P2.
    \item At $t=t_0$, we trace a magnetic field line from P2 and record the connectivity (O, C, or D).
    \item We advect the end-point of the field line forward in time from $t_0$ to $t_1$ using
    $\mathbf{\tilde{v}}(t)$. We label the final point P3.
    \item At $t=t_1$, we trace a magnetic field line from P3 and record the connectivity (O, C, or D).
\end{enumerate}
Notice that P2 and P3 may be a little above or below but still close to the associated slip surface.

Analogously, we obtain the dual slip-back mapping (Fig.~\ref{fig:mapping}b) through
the following steps
\begin{enumerate}
    \item At $t=t_1$, we trace a magnetic field line from the initial point P1 
    and record the connectivity (O, C, or D) as in the primary mapping.
    \item P1 is then advected back in time from $t_1$ to $t_0$ with the  $\mathbf{\tilde{v}}(t)$ flow. We label the final point D2.
    \item At $t=t_0$, we trace a magnetic field line from D2  and record the connectivity (O, C, or D).
    \item We advect the end-point of the field line forward in time from $t_0$ to $t_1$ using
    $\mathbf{\tilde{v}}(t)$. We label the final D3.
    \item At $t=t_1$, we trace a magnetic field line from D3 and record the connectivity (O, C, or D).
\end{enumerate}
Again, D2 and D3 are generally close but removed from the associated slip surface.
We now have all the elements to obtain the codes described in \S \ref{sec:sbm}.
\end{document}